# N-Doped Surfaces of Superconducting Niobium Cavities as a Disordered Composite – addendum

Wolfgang Weingarten[1], formerly CERN now retired
Esplanade des Particules 1
P.O. Box
1211 Geneva 23
Switzerland

**Abstract**
Experimental data on N-doped superconducting (sc) cavities are further analyzed which contribute to confirm a recently proposed model explaining the increase of the Q-value versus the accelerating gradient.

The present analysis is based on several publications. These include the proposed model [1], data from Cornell University [2], a conference paper [3], a study on the production and purification of niobium [4], and another study on dissolved gases in niobium [5].

However, this model was criticized in ref. 3. It was stated to be in direct contradiction with the experimental observations as published in ref. 2, and therefore not suitable to describe increase of the Q-value versus the accelerating gradient (positive Q-slope). The cause of the criticism is a linear relationship in ref. 1 between the mean free path l and the RRR value, which is valid for l of some size larger than the coherence length ξ. but this was not mentioned in ref. 1. For small l ≈ ξ, the linear relationship breaks down. However, the authors of ref. 3 applied this relationship for small l as well, resulting in inconsistencies with the data.

The criticism of ref. 3 has already been objected to, but only in a short note in another context [6]. Therefore, a more detailed analysis is in order, which will be given below.

The above-mentioned experimental observations show the surface resistance of accelerating cavities at 1.3 GHz made of niobium sheet "doped" with nitrogen [7]. The data cover the temperature interval from 1.5 to 2.11 K. Three doping treatments changed the mean free path length (4.5 nm, 34 nm, and 213 nm) and indeed showed a decrease of the surface resistance with the RF field, except for the last treatment.

In the following these data are analysed in several steps.

In the <u>first step</u>, at low field strength, the temperature-dependent (and field-independent) part of the surface resistance $R_s$ (usually called "BCS surface resistance" $R_{BCS}$) is studied as a function of temperature T (Fig. 1), corresponding to $R_{BCS} \sim e^{-\Delta/T}/T$, where $\Delta$ is the energy gap.

It turns out that the semilogarithmic plot shows a straight line, suggesting superconductivity. However, the slope is slightly larger than the commonly observed energy gap $\Delta = 18.9$ K, about $\Delta = 20$ K. This observation in turn leads to a critical temperature $T_c$ of 9.5 K.

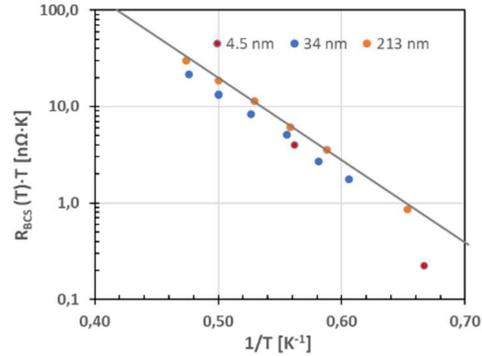

Fig. 1: Dependence of the "BCS-surface resistance" $R_{BCS}$ on the bath temperature T.

In the <u>2nd step</u>, the surface resistance $R_{BCS}$ at low field is represented by matching it to the expression:

$$R_{BCS}(\omega, T) = \underbrace{\frac{1}{2} \cdot \mu_0^2 \omega^2 \lambda^3 s_{Nb}}_{A} \cdot \overbrace{\frac{\Delta}{k_B T} \ln\left(\frac{\Delta}{\hbar\omega}\right) e^{-\Delta/T}}^{t(T)}. \quad (1)$$

Table 1: Fit parameter in relation with Fig. 2

| | | |
|---|---|---|
| Angular frequency | $\omega$ [GHz] | 18.8 |
| Mean free path | $l$ [nm] | 2.7·RRR |
| Coherence length | $\xi_0$ [nm] | 39 |
| Penetration depth | $\lambda_L$ [nm] | 38 |
| Electrical conductivity | $s_{Nb}$ [(Ωm)$^{-1}$] | 3.0·10$^8$ |
| Residual resistivity ratio | RRR | 40 |
| Energy gap | $\Delta/k_B$ [K] | 17.8 |
| $s_{Nb}=s_{Nb}$(300 K)·RRR; $s_{Nb}$(300 K) = 7.6·10$^6$ (Ωm)$^{-1}$ | | |

---

[1] wolfgangweingarten@t-online.de

Eq. 1 is derived from the modified two-fluid surface resistance and needs a numerical justification. This is provided by using surface resistance data as a function of temperature with known electrical conductivity at low temperature or the RRR-value respectively [8], Fig. 2.

Eq. 1 represents well the data of Fig. 2. The relevant parameters are listed in Table 1, with the aid of

$$\frac{1}{\xi} = \frac{1}{\xi_0} + \frac{1}{l} \quad \text{and}$$

$$\lambda = \lambda_L \sqrt{1 + \frac{\xi}{l}} \quad .$$

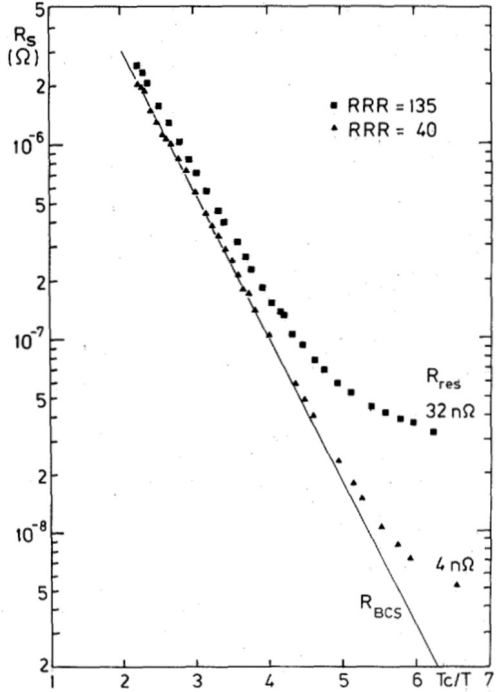

Fig. 2: Surface resistance of a 3 GHz niobium sheet metal cavity with known RRR value (adopted from ref. 8). The fit of $R_{BCS}$ using eq. 1 under the parameters of Table 1 coincides with the straight line.

A field-dependent part of $R_s$ must be taken account of, which can be parameterized as follows,

$$R_s(\omega, T, B) = R_{BCS}(\omega, T) \cdot h(B), \quad (2.a)$$

$$h(B) = 1 - f(B) + cf(B), \quad (2.b)$$

$$f(B) = \frac{\ln(B/B^*)}{\ln(B_c^*/B^*)} . \quad (2.c)$$

B is the RF surface magnetic field, $B^*$ is the initial field of the Q-slope (~10 mT) and $B_c^*$ is the saturation field of the Q-slope (~93 mT). The full amplitude of the magnetic field dependent part determines the constant c (Table 2), c = $s_m/s_{Nb}$ (symbols to be explained later).

Since the parameter $\Delta$ is known from Fig. 1 and $B^*$, $B_c^*$ are known from inspection of the data (Fig. 3), the only free fitting parameters with respect to eqs. 1 and 2 are A and c. The obvious fact, which is confirmed in Table 2, namely that c is fairly independent of T, justifies the factorization as in Eq. 2.a.

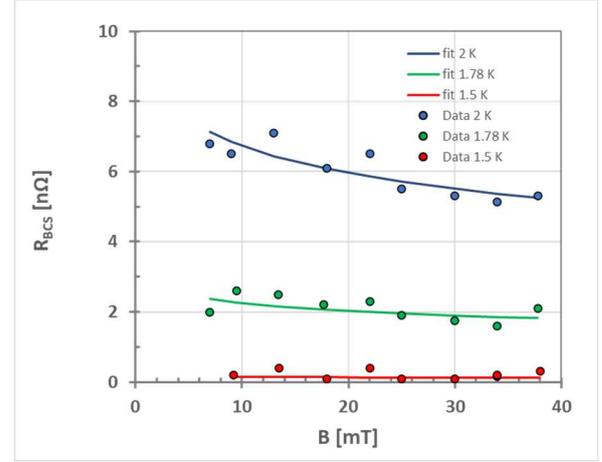

(a)

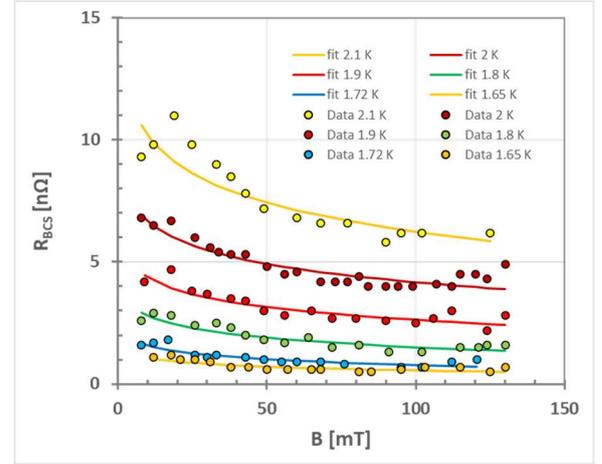

(b)

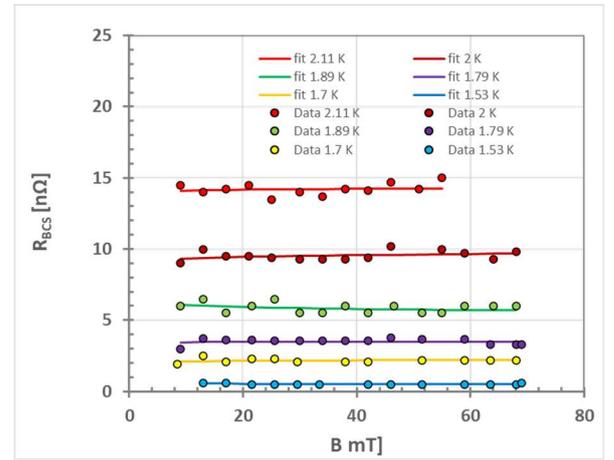

(c)

Fig. 3: Surface resistance $R_{BCS}$ vs. peak magnetic surface field B for 1.3 GHz niobium sheet cavity at different temperatures. The dots represent the data measured at Cornell University [2], the lines indicate the fitting results with the parameters as in Table 2. The three plots were obtained for three different mean free paths l = 4.5, 34 and 213 nm (from a to c) 2.

In the 3rd step, Landauer's percolation model is used [9], modified as in ref. 1, eq. 3. It describes the overall electrical conductivity $s_m$ of a binary mixture of component 1 (electrical conductivity $s_1$) and component 2 (electrical conductivity $s_2$).

Component 1 here represents the "weak" superconductor, which is subject to the proximity effect of component 2, the "strong" superconductor. $x_1$ is the fraction of the total volume occupied by the "weak" superconductor, and $x_2$ represents the fraction of the total volume occupied by the strong superconductor, $x_2 = 1 - x_1$. The constant $c = s_m/s_{Nb}$ describes the decrease (or increase) of the overall electrical conductivity $s_m$ under the exposure of the magnetic field B related to the low field electrical conductivity $s_{Nb}$ (when the weak superconductor is still sc and has not yet started changing to the normal conducting (nc) state).

Table 2: Fitting parameters A and c related to data of Fig. 3 [1)]

| l [nm] | T [K] | A [nΩ] | c | l [nm] | T [K] | A [nΩ] | c | l [nm] | T [K] | A [nΩ] | c |
|---|---|---|---|---|---|---|---|---|---|---|---|
| 4.5 | 2 | 5103 | 0.64 | 34 | 1.9 | 5439 | 0.60 | 213 | 2 | 7118 | 1.04 |
| 4.5 | 1.78 | 5253 | 0.69 | 34 | 1.8 | 5877 | 0.55 | 213 | 1.89 | 7749 | 0.95 |
| 4.5 | 1.5 | 2379 | 0.71 | 34 | 1.72 | 5278 | 0.50 | 213 | 1.79 | 7542 | 1.04 |
| 34 | 2.1 | 5149 | 0.61 | 34 | 1.65 | 5588 | 0.54 | 213 | 1.7 | 8000 | 1.05 |
| 34 | 2 | 5053 | 0.64 | 213 | 2.11 | 6736 | 1.02 | 213 | 1.53 | 7009 | 0.88 |

[1)] error intervals for A and c are about ± 10%

The "weak" superconductor remains sc due to the proximity effect up to a small critical field B* (onset field), then it transitions to the nc state. While the rf field B continues to increase, a larger and larger fraction of the weak superconductor becomes nc and penetrates deeper into the surface until the weak superconductor is exhausted at a distance from the surface that defines the saturation field $B_c^*$. The second component of the binary mixture, the "strong" superconductor, consists of relatively pure niobium metal. Its electrical conductivity $s_2$ is purely imaginary,

$$s_2 = \frac{i}{\mu_0 \lambda_L^2 \left[1 + \frac{1}{1 + l/\xi_0}\right] \omega} \qquad , \qquad (3)$$

with $\lambda_L$ the London penetration depth (3.8·10⁻⁸ m), $\xi_0$ the coherence length (3.9·10⁻⁸ m), and $\omega$ the angular frequency (8.17·10⁹ s⁻¹).

The modified Landauer model shows a uniform increase in $s_m$ with $x_1$, culminating in a percolation maximum near $x_1 = 0.67$, and a uniform decrease thereafter. This maximum is considered to be the dominant contribution to the surface resistance.

In the reverse order, $s_m$ can now be determined: $s_m = c \cdot s_{Nb}$, because c and $s_{Nb}$ are known from fitting to the data (Table 3).

Table 3: Calculation of the electrical conductivities $s_m$ and $s_1$

| l [nm] | c (average) | $s_{Nb}$ [(Ωm)⁻¹] | $s_m$ [(Ωm)⁻¹] | $s_1$ [(Ωm)⁻¹] | $\rho_1$ [μΩcm] |
|---|---|---|---|---|---|
| 4.5 | 0.68 ± 0.03 | (3.1 ± 1.4)·10⁸ | 2.09·10⁸ | 4.85·10⁶ | **21 ± 15** |
| 34 | 0.57 ± 0.05 | (4.9 ± 0.3)·10⁸ | 2.83·10⁸ | 7.19·10⁶ | **14 ± 3** |
| 213 | 1.00 ± 0.04 | (1.03 ± 0.06)·10⁹ | 1.03·10⁹ | 7.02·10⁷ | **1.4 ± 0.2** |

All elements are now available to determine $s_1$ at the percolation maximum ($x_1 = 0.67$) and the corresponding electrical resistivity $\rho_1$ by using the formula in ref. 1, eq. 3 (last two columns of Table 3).

In the 4th step, the result for $\rho_1$ is to be compared with data known from elsewhere. This comparison shall finally provide a criterion for the validity of the presented model.

Table 4: Determination of the electrical resistivity $\rho_1$ of "weak" component [2)]

| l [nm] | RRR | wt. % N | at. % N | $\rho_1$ [μΩcm] |
|---|---|---|---|---|
| 4.5 | 0.64 | 0.61 | 4.0 | **23** |
| 34 | 0.95 | 0.41 | 2.7 | **16** |
| 213 | 9,24 | 0.04 | 0.3 | **1.6** |

[2)] The RRR of the "weak" component is defined as RRR = $s_1/s_{Nb}$ (300 K); $s_{Nb}$ (300 K) = 7.6·10⁶ [(Ωm)⁻¹]; $s_1$ from Table 3.

DeSorbo finds for 0.23 (0.33, 1.64) at. % nitrogen interstitially dissolved in niobium a low temperature electrical resistivity of 1.7 (1.9, 1.8) μΩcm. Padamsee gives an RRR value of 3900 for 1 wt. ppm nitrogen. In Table 4, the electrical resistivity $\rho_1$ is determined from the RRR value. Within error margins, equivalence is found between $\rho_1$ from Table 3 and Table 4 (bold numbers in red). Thus, the model is internally consistent

In summary, the model presented in ref. 1 enables the understanding of the data in Fig. 3. The claim from ref. 3 that the model is in direct contradiction with the experimental observations is therefore rejected.

# References

[1] W. Weingarten, N-doped surfaces of superconducting niobium cavities as a disordered composite, IEEE Trans. Appl. Supercond. **28**, 3500504 (2018).

[2] J. T. Maniscalco, D. Gonnella, and M. Liepe, The importance of the electron mean free path for superconducting radio-frequency cavities, J. Appl. Phys. **121**, 043910 (2017).

[3] J. T. Maniscalco, M. Ge, P. N. Koufalis, M. Liepe, T. A. Arias, D. B. Liarte, J. P. Sethna, and N. Sitaraman, Proc. 19th Int. Conf. on RF Superconductivity, SRF2019, Dresden, Germany, 340.

[4] H. Padamsee, The technology of Nb production and purification, Proceedings of SRF Workshop 1984, Geneva, Switzerland, p. 339.

[5] W. DeSorbo, Effect of dissolved gases on some superconducting properties of niobium, Phys. Rev. **132**, 107 (1963).

[6] W. Weingarten, Fluxon-induced losses in niobium thin-film cavities revisited, IEEE Trans. Appl. Supercond. **30**, 3500411 (2020).

[7] A. Grassellino, A. Romanenko, D. Sergatskov, O. Melnychuk, Y. Trenikhina, A. Crawford, A. Rowe, M. Wong, T. Khabiboulline and F. Barkov, Nitrogen and argon doping of niobium for superconducting radio frequency cavities: a pathway to highly efficient accelerating structures, Supercond. Sci. Technol. **26,** 102001 (2013).

[8] H. Lengeler, W. Weingarten, G. Müller, H. Piel, Superconducting niobium cavities of improved thermal conductivity, IEEE Trans. Magnet. **MAG-21**, 1014 (1985).

[9] R. Landauer, The electrical resistance of binary metallic mixtures, J. Appl. Phys. **23**, 779 (1952).